\documentclass[11pt]{article}
\usepackage{amsmath}
\usepackage{jheppub}
\usepackage{amssymb}
\usepackage{graphicx}
\usepackage[T1]{fontenc}
\usepackage{slashed}
\usepackage{hyperref}
\usepackage[utf8]{inputenc}
\usepackage{xspace}
\usepackage{color}
\usepackage{ulem}
\usepackage{array}
\usepackage{ulem,fancyvrb}
\usepackage{xcolor}
\usepackage{mathtools}
\date{\today}

\begin{document}
%\linenumbers

\title{Three channel dissipative warm Higgs inflation with global inference via genetic algorithms}

\author[a,b]{Wei Cheng}
\affiliation[a]{School of Science, Chongqing University of Posts and Telecommunications, Chongqing 400065, China}
\affiliation[b]{Department of Physics and Chongqing Key Laboratory for Strongly Coupled Physics, Chongqing University, Chongqing 401331, China}
\emailAdd{chengwei@cqupt.edu.cn}

\abstract{This paper constructs and analyzes a three channel dissipative framework for Warm Higgs Inflation, wherein the total dissipation coefficient, $\Upsilon(h,T)$, is decomposed into low temperature, high temperature, and threshold activated contributions. A genetic algorithm is employed for the global numerical solution and statistical inference of the background field dynamics. To overcome the single channel dominance degeneracy in high dimensional parameter scans, two classes of structural priors are introduced into the objective function: a \texttt{mixing} prior to suppress extreme channel fractions and an \texttt{entropy} prior to favor multi channel coexistence. Furthermore, the adoption of a layered warmness criterion (e.g., $Q \equiv \Upsilon/3H$) decouples model selection from cosmological observables, thereby enhancing analytical transparency.

The complete workflow is demonstrated on a $14$ dimensional phenomenological model. An ablation study of the priors (\texttt{noprior} vs. \texttt{mixing} vs. \texttt{mixing+entropy}) yields $18871$ viable parameter points, revealing that the priors significantly enhance the discovery probability of non-trivial multi channel solutions within a parameter space naturally biased towards single channel dominance. After imposing the observational constraint $r < 0.036$, the number of retained solutions for each scenario is $14485$, $1889$, and $1971$, respectively. A typical best fit solution exhibits a "channel relay" dynamical feature during its evolution and a genuinely mixed state at the pivot scale (e.g., $f_{{\rm HT},*}\simeq 0.399480$, $f_{{\rm th},*}\simeq 0.600517$), implying that the microscopic origin of dissipation need not be unique within a single inflationary history. The energy conservation residual is stably controlled at the level of $\mathcal{O}(10^{-7} \to 10^{-4})$, which ensures the numerical reliability of the background evolution and establishes a foundation for further cosmological perturbation spectra calculations.
}

\maketitle
\preprint{}

\section{Introduction}
The theory of cosmic inflation provides a dynamical attractor solution for the early universe, rendering its evolution largely insensitive to initial conditions. It concurrently resolves the horizon and flatness problems while accounting for the near scale-invariance of primordial perturbations~\cite{Guth:1980zm,Linde:1981mu,Starobinsky:1980te,Mukhanov:1981xt,Sasaki:1986hm,Lyth:2009imm}. In the standard cold inflation, any pre-existing radiation is exponentially diluted, necessitating a subsequent reheating phase to re-establish the universe's thermal history. Warm Inflation (WI) merges these two epochs by positing that dissipative interactions continuously source a thermal radiation bath from the inflaton's energy density~\cite{Berera:1995wh,Berera:1996nv,Berera:1996fm,Berera:1998gx,Freese:2024ogj,Rodrigues:2025neh,Berghaus:2025dqi}. This process maintains a non-zero temperature ($T>0$) and introduces a crucial feedback mechanism from the radiation bath into the slow-roll dynamics. This not only modifies the background evolution but also fundamentally alters the noise sources and transfer functions governing the generation of perturbations. Consequently, when confronted with observational data, WI carves out a distinct parameter space from its cold inflation counterpart~\cite{Arya:2017zlb,Panotopoulos:2015qwa,Bastero-Gil:2015nja,Visinelli:2011jy,Reyimuaji:2020bkm,Rasouli:2018kvy}.

Over the past decade, research in WI has advanced along three principal fronts: (i) Microscopic Origins: constructing well controlled dissipative coefficients, $\Upsilon$, from first principles in finite-temperature field theory and scrutinizing the self-consistency of thermalization and reaction rates~\cite{Berera:1999ut,Bastero-Gil:2010dgy}, (ii) Perturbation Theory: investigating how the interplay of thermal noise and dissipative fluctuations impacts predictions for the scalar spectral index ($n_s$) and the tensor-to-scalar ratio ($r$)~\cite{Moss:2007cv,Taylor:2000ze,Moss:2008yb,Moss:2011qc,Bastero-Gil:2011rva,Ramos:2013nsa}, and (iii) Model Building: developing a diverse array of models, from the Warm Little Inflaton to Warm Higgs Inflation (WHI), and confronting them with observational constraints from Planck, BICEP/Keck, and other experiments~\cite{Montefalcone:2022jfw,Bastero-Gil:2016qru,Eadkhong:2023ozb,Planck:2018jri,BICEP:2021xfz}.

These collective advancements highlight a formidable challenge: dissipation is not a monolithic phenomenon describable by a single functional form. Its effective description is highly sensitive to the temperature regime, particle spectrum, and coupling structure, with the dominant mechanism potentially transitioning during the inflationary evolution~\cite{Berera:1995ie,Berera:2008ar,Bastero-Gil:2006ahd,Bastero-Gil:2012akf}. A common practice in phenomenological studies is to adopt a single channel, power law parameterization (e.g., $\Upsilon\propto T^n\phi^m$) and perform local scans or Markov Chain Monte Carlo (MCMC) analyses~\cite{Bartrum:2013oka,Bastero-Gil:2017wwl,Kumar:2024hju,Santos:2024pix}. However, if the true dissipation arises from a superposition of multiple sources, this single channel assumption forcibly projects the competition and succession of different mechanisms onto a single effective curve, rendering the underlying physics indistinguishable~\cite{Graham:2009bf,Hall:2003zp,Bartrum:2013fia,Benetti:2016jhf,Cheng:2024uvn}. This issue, often debated as "scheme-dependence", underscores a structural uncertainty that transcends mere numerical precision. Motivated by this, $\Upsilon(h,T)$ is treated as a decomposable object, and its channel structure is inferred directly from the underlying dynamics~\cite{Berera:1998px,Berera:2008ar,Bastero-Gil:2012akf}.

More critically, global search algorithms in high dimensional spaces particularly those involving stiff ODEs and parameters spanning orders of magnitude exhibit an intrinsic bias~\cite{Hairer,Gabor,Donoho}. They naturally favor extremal solutions, pushing weights towards a single channel to most rapidly minimize the objective function~\cite{Roberts,Daniel}. If not explicitly counteracted, this algorithmic bias can be misconstrued as a physical preference~\cite{Trotta:2008qt,Betancourt:2017ebh}. This phenomenon is termed "dominant channel degeneracy". This statistical bias is robust, appearing consistently across independent algorithms (e.g., DE, MCMC, grid searches) even without explicit priors~\cite{Storn,JMLR:v13:bergstra12a}.

The objective of this paper is therefore not merely to find best fit points for a fixed $\Upsilon$ form~\cite{Bastero-Gil:2017wwl,Benetti:2016jhf,Kumar:2024hju,Santos:2024pix,Eadkhong:2023ozb,Arya:2018sgw}, but to construct a \textit{diagnosable inference framework}. This framework is designed to: (1) render multi channel dynamics an explicit numerical observable by outputting channel fractions $f_i(N)$ during integration; (2) suppress degenerate attractors during inference by incorporating structural priors that regularize the maximum channel fraction ($f_{\max,*}$) and entropy ($H_n$), thereby yielding a statistically interpretable sample of mechanisms; and (3) provide transparent and traceable selection effects by decoupling the primary scan from the strict warmness screening. To achieve these goals, a closed-loop workflow is implemented, encompassing theoretical model construction, numerical simulation, global inference via a Genetic Algorithm (GA), and finally, statistical analysis and observational verification.

The remainder of this paper is organized as follows.
In Sec.~\ref{sec:WHIBE}, the theoretical framework for WHI in the Einstein frame is established, with the effective potential and background equations of motion elaborated in detail.
Sec.~\ref{sec:TCDCP} introduces a unified parameterization for the low temperature (LT), high temperature (HT), and threshold activated (th) dissipation channels and defines their respective fractional contributions.
In Sec.~\ref{sec:NI}, the procedure for extracting observables at the pivot scale is described, and the self-consistency checks for the inflationary solution are outlined.
Sec.~\ref{sec:GI} presents the objective function for our GA based global inference, with a focus on the formulation and role of the mixing and entropy structural priors.
The main numerical results are presented and physically interpreted in Sec.~\ref{sec:RD}.
Finally, a summary of the key findings is provided, and conclusions are drawn in Sec.~\ref{sec:Sum}.

\section{WHI: theoretical framework and background dynamics}\label{sec:WHIBE}

This section establishes the minimal theoretical framework for the analysis. First, the effective potential for WHI in the Einstein frame, $V_E(h)$, is presented, along with the coupled system of background equations governing the warm inflation dynamics. These equations form the basis of the numerical simulations and provide the physical constraints for the objective function used in the GA (Sec.~\ref{sec:GI}). Subsequently, the criteria for the termination of inflation are defined, and a metric for quantifying the energy conservation error is introduced. This lays the necessary groundwork for the numerical implementation and the stability analysis presented in Sec.~\ref{sec:RD}.

\subsection{Effective WHI potential in the Einstein frame}\label{sec:EP}

In the Einstein frame, the effective potential for WHI with a non-minimal coupling to gravity is given by~\cite{Cheng:2024uvn,Kallosh:2013hoa}:
\begin{equation}
V_E(h) \simeq \frac{\lambda M_p^4}{4\xi^2}\left[1+\exp\left({-\sqrt{\frac{2}{3}}\frac{h}{M_p}}\right)\right]^{-2}.
\end{equation}
where $\lambda$ is the Higgs self-coupling constant and $\xi$ is the non-minimal coupling parameter. In the context of this work, $\lambda$ and $\xi$, in conjunction with the dissipative structure, collectively determine the evolution of the Hubble parameter $H$, the slow-roll parameter $\epsilon_H$, and ultimately, the cosmological observables.

\subsection{Background equations for WHI}\label{sec:WBE}

The background dynamics of warm inflation are described by a coupled system comprising the inflaton's equation of motion, the evolution equation for the radiation energy density, and the Friedmann constraint~\cite{Berera:1995ie,Berera:2008ar,Baumann:2009ds,Vilenkin:2000jqa}. The explicit equations are:
\begin{align}
  \ddot{h} + (3H+\Upsilon)\dot{h} + \frac{dV_E}{dh} &= 0, \label{eq:inflaton_eom} \\
  \dot{\rho}_R + 4H\rho_R &= \Upsilon \dot{h}^2, \label{eq:radiation_eom} \\
  3M_P^2 H^2 &= \frac{1}{2}\dot{h}^2 + V_E(h) + \rho_R. \label{eq:friedmann}
\end{align}
Equation~\eqref{eq:inflaton_eom} describes the inflaton's motion under the combined influence of Hubble friction ($3H$) and dissipative friction ($\Upsilon$). Equation~\eqref{eq:radiation_eom} shows that the radiation bath is simultaneously diluted by cosmic expansion and replenished by the dissipative transfer of energy from the inflaton field. Equation~\eqref{eq:friedmann} enforces the total energy conservation of the system. The dissipation coefficient, $\Upsilon$, is the central element of "warmth", as it both enhances friction and sources the radiation bath~\cite{Taylor:2000ze,Berera:2008ar,Bartrum:2013fia}. Given that $\Upsilon$ can contain complex power law or exponential dependencies~\cite{Berera:2023liv,Bastero-Gil:2010dgy,Bastero-Gil:2012akf,Graham:2009bf,Hall:2003zp}, the numerical integration of this system is sensitive to step size control and truncation errors. To ensure the reliability of the results, both the satisfaction of termination conditions and the control of energy conservation errors are monitored throughout the simulations.

Key diagnostic quantities are defined as follows:
\begin{align}
  \rho_R &= \frac{\pi^2}{30}g_* T^4, \\
  Q &\equiv \frac{\Upsilon}{3H}, \\
  \epsilon_H &\equiv -\frac{\dot{H}}{H^2},
\end{align}
where $g_*$ is the effective number of relativistic degrees of freedom, the ratio $Q$ quantifies the strength of dissipation relative to Hubble damping, and $\epsilon_H$ is the primary slow-roll parameter. The condition $\epsilon_H \ll 1$ signifies a period of accelerated expansion, while $\epsilon_H = 1$ is conventionally used to mark the end of inflation. In the present framework, the "warmness" indicators, such as $T/H$ and $Q$, are treated as hierarchical constraints that can be applied either as soft selections during the main scan or as hard cuts in post-processing.

\section{Three channel dissipation and fractional contributions}\label{sec:TCDCP}

The dynamics of warm inflation are fundamentally governed by the dissipation coefficient, $\Upsilon$. When multiple physical sources of dissipation are present, a single channel parameterization obscures information about "mechanism competition" and can degrade the performance of statistical inference. This section, therefore, proposes a three channel decomposition of $\Upsilon$, with the fractional channel contribution, $f_i(N)$, serving as a primary output quantity. This choice directly facilitates the definition of the structural priors in Sec.~\ref{sec:GI}, as these priors will act not on abstract model parameters but on diagnosable, physical, structural quantities.

\subsection{Three channel decomposition and channel weights}

The three channel decomposition for the dissipation coefficient in WHI is defined as:
\begin{equation}
  \Upsilon(h,T) = \Upsilon_{\text{LT}}(h,T) + \Upsilon_{\text{HT}}(h,T) + \Upsilon_{\text{th}}(h,T),
\end{equation}
where $\Upsilon_{\text{LT}}$, $\Upsilon_{\text{HT}}$, and $\Upsilon_{\text{th}}$ denote the LT, HT, and th dissipation channels, respectively. The fractional contribution of each channel is then defined as:
\begin{equation}
  f_i(N) \equiv \frac{\Upsilon_i}{\Upsilon_{\text{LT}} + \Upsilon_{\text{HT}} + \Upsilon_{\text{th}}}, \quad \text{with} \quad \sum_i f_i = 1, \quad i \in \{\text{LT, HT, th}\}.
\end{equation}
The quantity $f_i$ directly quantifies the notion of "dominance". A state where $f_{\max} \to 1$ indicates single channel dominance and thus inference degradation, whereas multiple $f_i$ values of a similar order of magnitude signify the coexistence or relaying of mechanisms. It is emphasized that the channel share is a structural quantity, not a direct assertion of the physical existence of specific channels. It serves as an inferential tool to prevent model degradation and to provide statistical clues for subsequent mapping to microscopic theories.

\subsection{Unified parameterization of LT/HT channels}

To enable a global inference that is not strongly tied to specific microphysical models, a flexible interface of "amplitude $\times$ power law $\times$ shape function" is adopted. The amplitude is controlled by a coefficient $C_i$, the power law dependence by exponents such as $(\alpha_h, \alpha_T)$, and a replaceable shape function $S_i$ allows for the future embedding of more refined microscopic calculations. Specifically, this unified and interchangeable interface for the LT/HT channels are expressed as:
\begin{align}
  \Upsilon_{\text{LT}}(h,T) &= C_{\text{LT}}\,T^{p_{\text{LT}}}\,S_{\text{LT}}(h,T;\theta_{\text{LT}}), \\
  \Upsilon_{\text{HT}}(h,T) &= C_{\text{HT}}\,T^{p_{\text{HT}}}\,S_{\text{HT}}(h,T;\theta_{\text{HT}}).
\end{align}
Those formulation separates the overall magnitude ($C_i$) from the temperature and field dependent shape ($S_i$ and exponents), which facilitates the application of structural priors to the fractional contributions $f_i$ rather than to specific microphysical coupling constants.

The numerical examples in this paper employ the following prototype instances for $S_{\text{LT/HT}}$. A numerically stabilized field amplitude, $h_{\text{bar}} \equiv \sqrt{h^2+h_0^2}$ with $h_0>0$, is introduced to prevent pseudo-singularities as $h\to 0$.
\begin{itemize}
  \item \textbf{LT Prototype:} A phenomenological implementation of the low temperature $T^3$ behavior is given by the following equation, reflecting the typical distributed mass realization in the LT regime \cite{Bastero-Gil:2010dgy,Bastero-Gil:2012akf,Bastero-Gil:2018yen}:
    \begin{equation}
      \Upsilon_{\text{LT}}(h,T) = C_{\text{LT}}\,\frac{T^3}{h_{\text{bar}}^2}.
    \end{equation}
    Here, the $T^3$ dependence captures the characteristic phase space enhancement at low temperatures~\cite{Bastero-Gil:2012akf,Berera:2008ar}, while the denominator reflects the suppression of dissipation at large field values due to effective mass effects.
  \item \textbf{HT Prototype:} A general power law form is used to model behavior across a broader temperature range~\cite{Berera:2008ar,Bastero-Gil:2016qru}:
    \begin{equation}
      \Upsilon_{\text{HT}}(h,T) = C_{\text{HT}}\,T\left(\frac{h_{\text{bar}}}{\mu}\right)^{\alpha_h}\left(\frac{T}{\mu}\right)^{\alpha_T}.
    \end{equation}
    Here, $\mu$ is a reference scale ensuring the arguments of the power laws are dimensionless. The exponents $\alpha_h$ and $\alpha_T$ are treated as free parameters to be determined by the GA~\cite{Benetti:2016jhf}.
\end{itemize}

\subsection{threshold activated channel}

The threshold channel is designed to model the exponential switching behavior that arises from the competition between the effective mass of an intermediate state and the ambient temperature. This channel is Boltzmann-suppressed when $M_{\text{eff}}/T$ is large but can open rapidly and even dominate for a period as $M_{\text{eff}}/T$ decreases~\cite{Berera:2008ar,Cerezo:2012ub,Bartrum:2013fia}. Its parameterization is defined as:
\begin{equation}
  \Upsilon_{\text{th}}(h,T) = C_{\text{th}}\,\mu\left(\frac{|h|}{\mu}\right)^{a}\left(\frac{T}{\mu}\right)^{c}\exp\!\left[-\frac{M_{\text{eff}}(h,T)}{T}\right],
\end{equation}
with the effective mass given by:
\begin{equation}
  M_{\text{eff}}^2(h,T) = (g|h|)^2 + m_0^2 + \kappa_T T^2,
\end{equation}
where $g|h|$ is the field dependent mass, $m_0$ is a vacuum mass, and $\kappa_T T^2$ is the thermal mass correction~\cite{Dolan:1973qd,Landsman:1986uw,Quiros:1999jp,Laine:2016hma}. The exponential factor governs the "activation window" of this channel, making it crucial for realizing "channel relay" phenomenology.

\section{Numerical implementation and diagnostics}\label{sec:NI}

Having established the theoretical inputs, this section details the numerical protocol for the simulations. It specifies the use of the e-fold number as the independent variable, the definition of the inflation termination event, the procedure for identifying the pivot scale, and the method for extracting key quantities such as $Q_*$, $(T/H)_*$, and $f_{i,*}$. Furthermore, the energy conservation error is incorporated as an output metric, serving as a critical quality control indicator to distinguish physically valid solutions from numerically spurious artifacts. These protocols directly inform the fitness evaluation in the GA (Sec.~\ref{sec:GI}) and the data visualization in Sec.~\ref{sec:RD}.

\subsection{Pivot scale and key diagnostic measures}

The dissipation ratio and the warmness indicator at the pivot scale are defined as:
\begin{equation}
  Q_* = \frac{\Upsilon_*}{3H_*}, \qquad \left(\frac{T}{H}\right)_* = \frac{T_*}{H_*}.
\end{equation}
Here, $Q_*$ measures the strength of dissipative friction relative to Hubble friction, while $(T/H)_*$ compares the thermal energy scale to the Hubble scale. Within the "layered warmness" strategy of this paper, these quantities can serve as soft penalty targets in the primary scan or as hard thresholds for post-processing to screen for warm sub-samples.

\subsection{Inflation termination and e-fold variable}

The condition $\epsilon_H = 1$ is used to trigger the termination of the inflationary epoch. The number of e-folds, $N$, serves as the independent variable for tracking the evolution of all background quantities. In graphical representations, the functions $\epsilon_H(N)$, $Q(N)$, $h(N)$, and $T/H(N)$ collectively form a visual chain of evidence, confirming that a period of inflation occurred and properly concluded.

\subsection{Energy conservation error}

The energy conservation error is defined by the fractional residual of the Friedmann equation:
\begin{equation}
  \delta_{\text{cons}}(t) \equiv \frac{\left|3M_P^2H^2 - \left(\frac{1}{2}\dot{h}^2 + V_E + \rho_R\right)\right|}{3M_P^2H^2}.
\end{equation}
The maximum value of this quantity over the entire evolution, $\text{cons\_err}_{\max} = \max_t \delta_{\text{cons}}(t)$, is used as a key diagnostic. A spike in this error at a critical evolutionary point typically indicates a need to adjust the numerical integration strategy. In this work, a penalty term, $\chi^2_{\text{cons}}$, derived from this error is incorporated into the total GA objective function to enforce numerical consistency among all viable samples.

\section{Global inference via GA with structural priors}\label{sec:GI}

This section details the "integrate-read-score" cycle of the GA and the total objective function. The mathematical definitions and physical motivations for the each objective function are elaborated. Those exposition provide the logical foundation for the "prior ablation" study and the comparative results presented in Sec.~\ref{sec:RD}.

\subsection{Parameter vector, encoding, and viability criteria}

The full parameter set subject to the scan is encapsulated in the vector:
\begin{equation}
\theta=\{\lambda,\xi,C_{\text{LT}},C_{\text{HT}},C_{\text{th}},g,a,c,\alpha_h,\alpha_T,h_0,\mu,m_0,\kappa_T\}.
\end{equation}
For the numerical implementation, multi scale positive definite parameters (e.g., $C_i, g, m_0$) employ a logarithmic encoding, while exponents are encoded linearly within physically motivated prior ranges. A parameter set (an "individual" in GA terminology) is deemed viable only if it satisfies a set of baseline criteria: the successful termination of inflation ($\epsilon_H \to 1$), a controlled energy conservation error, and the stable readout of finite values for the pivot-scale quantities ($Q_*$, $(T/H)_*$, $f_{i,*}$). Further details on parameter ranges and numerical settings are provided in Appendices A and B.

\subsection{Total objective function}
\label{subsec:total_objective}

A GA search is conducted by minimizing a composite objective (loss) function constructed as a sum of physically motivated and numerically stabilizing penalty terms. The total objective function is defined as
\begin{equation}
\chi^2_{\rm tot}
=\chi^2_{nsr}+\chi^2_{\rm warm}+\chi^2_{\rm mix}+\chi^2_{\rm ent}+\chi^2_{\rm each}+\chi^2_{N}+\chi^2_{\rm cons}.
\label{eq:chi2_tot_full}
\end{equation}

Each component serves a distinct purpose: $\chi^2_{nsr}$ constrains the solution to the observationally favored $(n_s,r)$ region, the term $\chi^2_{\text{warm}}$ applies a soft penalty for warmness conditions, the two terms, $\chi^2_{\text{mix}}$ and $\chi^2_{\text{ent}}$, form the core of the methodology presented here, designed to counteract single channel dominance and enable the identification of the underlying dissipation structure, $\chi^2_{\rm each}$ discourages solutions where any single channel fraction becomes vanishingly small, improving robustness against near-degenerate single-channel dominance, $\chi^2_{N}$ ensures the total e-folds $N_{\text{final}}$ falls within an expected range, $\chi^2_{\text{cons}}$ enforces numerical consistency by penalizing energy non-conservation.

For numerically infeasible candidates (e.g.\ failed integration, non-finite observables, or missing pivot diagnostics), a large constant penalty is assigned,
\begin{equation}
\chi^2_{\rm tot}=\chi^2_{\rm fail},
\label{eq:chi2_fail}
\end{equation}
ensuring such individuals have negligible selection probability. In addition, the default hyperparameters (as used in the implementation) are tabulated in Tab.~\ref{tab:hyperparams} of Appendix C.

Prior to elaborating on each component, the concept of Pivot-scale channel fractions is first introduced for the convenience of subsequent reference.
\paragraph{Pivot-scale channel fractions.}
The total dissipation is decomposed into three contributions at the pivot scale,
\begin{equation}
\Upsilon_*=\Upsilon_{{\rm LT},*}+\Upsilon_{{\rm HT},*}+\Upsilon_{{\rm th},*},
\label{eq:Gamma_decomp}
\end{equation}
and the corresponding channel fractions are defined as
\begin{equation}
f_{{\rm LT},*}\equiv \frac{\Upsilon_{{\rm LT},*}}{\Upsilon_*},
\qquad
f_{{\rm HT},*}\equiv \frac{\Upsilon_{{\rm HT},*}}{\Upsilon_*},
\qquad
f_{{\rm th},*}\equiv \frac{\Upsilon_{{\rm th},*}}{\Upsilon_*}.
\label{eq:f_def}
\end{equation}
The maximum channel fraction is
\begin{equation}
f_{\max,*}\equiv \max\{f_{{\rm LT},*},\,f_{{\rm HT},*},\,f_{{\rm th},*}\}.
\label{eq:fmax_def}
\end{equation}

\paragraph{(i) Observational term $\chi^2_{nsr}$.}
The primary observational guidance is imposed in the $(n_s,r)$ plane. A Gaussian penalty is adopted for the scalar spectral index, while the tensor-to-scalar ratio is controlled by an upper-bound penalty, reflecting the fact that current data predominantly constrain $r$ from above. The term is written as
\begin{equation}
\chi^2_{nsr}
=\left(\frac{n_s-n_s^{\rm obs}}{\sigma_{n_s}}\right)^2
+w_r\Bigl[\max\bigl(0,\,r-r_{\max}\bigr)\Bigr]^2,
\label{eq:chi2_nsr}
\end{equation}
where $n_s^{\rm obs}$ and $\sigma_{n_s}$ denote the reference central value and uncertainty used in the scan, $r_{\max}$ is the adopted upper limit, and $w_r$ controls the relative stiffness of the $r$-penalty.

The cosmological observables $n_s$ and $r$ take the following form:
\begin{align}
  n_s &\approx 1 - \frac{6\epsilon_V - 2\eta_V}{1+Q_*}, \\
  r &\approx \frac{16\epsilon_V}{(1+Q_*)^2},
\end{align}
where the potential based slow-roll parameters are defined as:
\begin{equation}
  \epsilon_V \equiv \frac{M_P^2}{2}\left(\frac{V_E'}{V_E}\right)^2, \qquad
  \eta_V \equiv M_P^2\frac{V_E''}{V_E}.
\end{equation}
In the cold inflation limit ($Q_* \to 0$), these expressions correctly reduce to the standard slow-roll formulae. For $Q_*>0$, the factor of $(1+Q_*)$ effectively suppresses the influence of the potential's shape on the spectral tilt and the tensor-to-scalar ratio.

\paragraph{(ii) Warmness term $\chi^2_{\rm warm}$ (soft penalty and optional hard cut).}
Warm inflation requires both a non-negligible dissipation strength and a radiation bath with temperature exceeding the Hubble scale. These properties are diagnosed at the pivot scale (denoted by ``$*$'') using
\begin{equation}
Q_*\equiv \frac{\Upsilon_*}{3H_*},
\qquad
\left(\frac{T}{H}\right)_*\equiv \frac{T_*}{H_*},
\label{eq:warm_defs}
\end{equation}
where $\Gamma_*$ is the total dissipation coefficient evaluated at the pivot scale. During the main scan, a soft warmness penalty is applied in logarithmic space to remain sensitive across multiple orders of magnitude:
\begin{align}
\chi^2_{\rm warm,soft}
&=
W_{T/H}(l)\Bigl[\max\!\bigl(0,\log_{10}(T/H)_{\min}-\log_{10}(T/H)_*\bigr)\Bigr]^2
\nonumber\\
&\quad+
W_{Q}(l)\Bigl[\max\!\bigl(0,\log_{10}Q_{\min}-\log_{10}Q_*\bigr)\Bigr]^2.
\label{eq:chi2_warm_soft}
\end{align}
where $(T/H)_{\min}$ and $Q_{\min}$ set the target warmness thresholds used for guidance. The weights $W_{T/H}(l)$ and $W_Q(l)$ may be annealed with generation index $l$ to reduce premature rejection of partially viable candidates at early stages while progressively enforcing warmness at late stages. The annealing prescription is
\begin{equation}
W_{T/H}(l)=A(l)\,W_{T/H}^{(0)},
\qquad
W_{Q}(l)=A(l)\,W_{Q}^{(0)},
\qquad
A(l)=A_{\rm floor}+\bigl(1-A_{\rm floor}\bigr)\left(\frac{l+1}{G}\right)^{p},
\label{eq:warm_anneal}
\end{equation}
where $G$ is the total number of generations, $A_{\rm floor}\in(0,1)$ prevents vanishing weights, and $p$ controls the rate of stiffening.

In addition, an optional hard warmness cut can be enabled inside the objective,
\begin{equation}
\chi^2_{\rm warm,hard}
=\chi^2_{\rm hard}\,\mathbf{\theta}\!\left[\left(\frac{T}{H}\right)_*<(T/H)_{\min}\ \ \text{or}\ \ Q_*<Q_{\min}\right],
\label{eq:chi2_warm_hard}
\end{equation}
where $\mathbf{\theta}[\cdots]$ denotes the indicator function:
\begin{equation}
\mathbf{\theta}[\mathcal{C}]=
\begin{cases}
1, & \mathcal{C}\ \text{true},\\
0, & \mathcal{C}\ \text{false}.
\end{cases}
\end{equation}

The warmness contribution used by the objective is then
\begin{equation}
\chi^2_{\rm warm}=
\begin{cases}
\chi^2_{\rm warm,hard}, & \text{hard-cut mode enabled},\\[4pt]
\chi^2_{\rm warm,soft}, & \text{soft-penalty mode}.
\end{cases}
\label{eq:chi2_warm_total}
\end{equation}
This layered design separates exploration (soft guidance) from strict selection (hard threshold), making selection effects explicit in post-processing.

The "warmness" of a solution is both a defining physical property and a potential source of selection effects. A layered strategy is therefore adopted. During the main scan, $\chi^2_{\text{warm}}$ acts as a soft penalty with an adjustable weight, preserving a diverse sample of mechanisms. Subsequently, hard thresholds (e.g., $T/H > 1$ and $Q > 10^{-3}$) are applied to the full data set in post-processing to extract a strictly warm sub sample. This approach makes the selection effects transparent by allowing for a direct comparison between the full and sub sampled distributions in the $n_s-r$ plane, rather than hiding these effects within the objective function's weighting scheme.

\paragraph{(iii) Mixing prior $\chi^2_{\rm mix}$.}
A common failure mode in high-dimensional dissipation scans is the emergence of spurious optima where one channel dominates while the remaining channels are effectively driven to zero. The mixing prior mitigates this pathology by (a) penalizing excessive dominance and (b) enforcing a minimal non-vanishing threshold-channel fraction. In the implementation-faithful form, define
\begin{equation}
\Delta_{\rm dom}\equiv \max\!\left(0,\frac{f_{\max,*}-f_{\rm dom}}{1-f_{\rm dom}}\right),
\qquad
\Delta_{\rm th}\equiv \max\!\left(0,\log_{10}f_{{\rm th,min}}-\log_{10}f_{{\rm th},*}\right),
\label{eq:mix_deltas}
\end{equation}
so that
\begin{equation}
\chi^2_{\rm mix}
=
W_{\rm dom}\,\Delta_{\rm dom}^{\,2}
+
W_{\rm th}\,\Delta_{\rm th}^{\,2}.
\label{eq:chi2_mix_full}
\end{equation}

To impose reasonable constraints on the dominance and threshold contribution in the algorithm, two key hyperparameters are defined: $f_{\text{dom}}$ serves to control the maximum allowable level of dominance (e.g., taking a value of 0.99), while $f_{\text{th},\text{min}}$ limits the minimum contribution of the threshold channel (e.g., no less than $10^{-3}$). Based on the characteristics of these parameters, two corresponding shift mechanisms are introduced: the normalized dominant shift $\Delta_{\text{dom}}$, which functions to maintain the comparability of penalty scales when $f_{\text{dom}} \to 1$; and the logarithmic threshold shift $\Delta_{\text{th}}$, which is designed to address scenarios where $f_{\text{th},*}$ decreases exponentially in such cases, linear penalties become numerically invalid, whereas $\Delta_{\text{th}}$ can provide robust sensitivity. The core objective of these designs ultimately lies in the constraint effect of $f_{\text{th},\text{min}}$: it can not only prevent the threshold channel from being erroneously "zeroed out" by the optimization algorithm, but also avoid the convergence of the GA to spurious optimal solutions at the boundaries of the parameter space, thereby restoring the physical interpretability of the results.

\paragraph{(iv) Entropy prior $\chi^2_{\rm ent}$.}
To complement the explicit dominance control, a softer global measure of multi channel sharing is introduced through the normalized information entropy of the channel fractions,
\begin{equation}
H_n \equiv -\frac{\sum_{i}\tilde f_{i,*}\ln \tilde f_{i,*}}{\ln 3},
\qquad
\tilde f_{i,*}\equiv \max(f_{i,*},\epsilon),
\qquad
i\in\{{\rm LT,HT,th}\},
\label{eq:Hn_def_full}
\end{equation}
with $\epsilon$ a small regulator preventing $\ln 0$. The penalty term is applied when $H_n$ falls below a minimum target value $H_{\min}$:
\begin{equation}
\chi^2_{\rm ent}
=
W_{\rm ent}\Bigl[\max\!\bigl(0,\,H_{\min}-H_n\bigr)\Bigr]^2.
\label{eq:chi2_ent_full}
\end{equation}
This prior suppresses sharply peaked channel distributions at the ensemble level without imposing strict lower bounds on each channel.

The normalized entropy $H_n$ approaches 0 for a pure single channel state and 1 for a uniform distribution~\cite{Shannon:1948,Thomas:2005}. This prior does not enforce lower bounds on individual channels but instead suppresses overly peaked distributions at the level of the entire solution cloud, fostering more stable multi channel statistics~\cite{Jaynes:1957zz,Gabriel}.

\paragraph{(v) Each-channel minimum prior $\chi^2_{\rm each}$.}
An additional implementation-level prior is included to explicitly discourage solutions where any individual fraction becomes numerically negligible. The term is defined as
\begin{equation}
\chi^2_{\rm each}
=
W_{\rm each}\sum_{i\in\{{\rm LT,HT,th}\}}
\left[\max\!\left(0,\log_{10}f_{{\rm each,min}}-\log_{10}f_{i,*}\right)\right]^2.
\label{eq:chi2_each_full}
\end{equation}
This component is particularly effective at preventing the GA from collapsing into near-degenerate manifolds corresponding to effectively reduced-channel dynamics.

\paragraph{(vi) e-fold control $\chi^2_{N}$.}
To avoid allocating excessive search effort to ultra-long inflationary trajectories, a mild penalty is applied when the total number of e-folds exceeds an upper reference bound:
\begin{equation}
\chi^2_{N}
=
w_N\Bigl[\max\bigl(0,\,N_{\rm final}-N_{\max}\bigr)\Bigr]^2.
\label{eq:chi2_N_full}
\end{equation}

\paragraph{(vii) Numerical-consistency term $\chi^2_{\rm cons}$.}
Numerical solutions are further filtered using an energy-conservation stability diagnostic. Let $\mathcal{R}_{E,\max}$ denote the maximal absolute value (over the integration interval) of the normalized residual of the continuity equation, computed from the discretized time series. A logarithmic penalty is imposed when $\mathcal{R}_{E,\max}$ exceeds a prescribed tolerance $\mathcal{R}_{E,{\rm thr}}$:
\begin{equation}
\chi^2_{\rm cons}
=
w_{\rm cons}\Bigl[\max\bigl(0,\log_{10}\mathcal{R}_{E,\max}-\log_{10}\mathcal{R}_{E,{\rm thr}}\bigr)\Bigr]^2.
\label{eq:chi2_cons_full}
\end{equation}
The use of a $\log_{10}$-distance yields stable discrimination across several orders of magnitude in the residual, which is essential when comparing solutions with substantially different stiffness properties.

\section{Results and discussion}\label{sec:RD}

This section presents the primary findings obtained from the GA scans. The analysis begins with an examination of representative solutions, illustrating the key dynamical features of multi channel WHI. Subsequently, a statistical overview of the results from the prior ablation study is provided, demonstrating the efficacy of the structural priors in mitigating single channel degeneracy. The physical implications of these findings are then discussed, establishing a clear connection between the numerical results, the underlying physical phenomena, and the constraints imposed by cosmological observations. This discussion serves to complete the argumentative loop, validating the inferential framework proposed in this work.

\subsection{Benchmark solution and Pivot-scale diagnostics}

A representative best fit solution, obtained from a run incorporating the structural priors, is presented to illustrate the key features of the framework. Tab.~\ref{tab:op} lists the parameter values for this benchmark point, while Tab.~\ref{tab:Piv} details the corresponding derived quantities at the pivot scale. This specific example serves to connect the subsequent analysis of the background evolution to a concrete point in the parameter space and to provide direct evidence that the structural priors indeed generate non-trivial multi channel solutions.

\begin{table}[htb]
\caption{Parameter values for a representative benchmark solution.}
\begin{center}
\begin{tabular}{c c c c}
\hline\hline
Parameter & Value & Parameter & Value \\
\hline
$\lambda$ & $1.0 \times 10^{-6}$ & $c$ & $-1.0$ \\
$\xi$ & $1.0$ & $\alpha_h$ & $0.7185$ \\
$C_{\text{LT}}$ & $16.1373$ & $\alpha_T$ & $-0.8$ \\
$C_{\text{HT}}$ & $3.4904$ & $h_0$ & $2.0280$ \\
$C_{\text{th}}$ & $1.2026$ & $\mu$ & $0.0613$ \\
$g$ & $1.0 \times 10^{-4}$ & $m_0$ & $1.0 \times 10^{-3}$ \\
$a$ & $2.9635$ & $\kappa_T$ & $15.8667$ \\
\hline\hline
\end{tabular}
\end{center}
\label{tab:op}
\end{table}

\begin{table}[htb]
\caption{Derived quantities and structural statistics at the pivot scale for the benchmark solution.}
\begin{center}
\begin{tabular}{c c c c}
\hline\hline
Quantity & Value & Quantity & Value \\
\hline
$n_s$ & $0.9551$ & $Q_*$ & $5.4 \times 10^{-3}$ \\
$r$ & $0.0061$ & $f_{\text{LT},*}$ & $3 \times 10^{-6}$ \\
$N_{\text{star}}$ & $11.3057$ & $f_{\text{HT},*}$ & $0.399480$ \\
$N_{\text{final}}$ & $61.3057$ & $f_{\text{th},*}$ & $0.600517$ \\
$(T/H)_*$ & $1.0121$ & $\text{cons\_err}_{\max}$ & $5.3 \times 10^{-3}$ \\
\hline\hline
\end{tabular}
\end{center}
\label{tab:Piv}
\end{table}

The data in Tab.~\ref{tab:Piv} reveal a significant contribution from the threshold channel at the pivot scale, with $f_{\text{th},*} \simeq 0.600517$. This indicates that the channel has been effectively activated. Concurrently, the HT channel contributes substantially, with $f_{\text{HT},*} \simeq 0.399480$, ruling out a scenario of complete single channel dominance. The resulting normalized entropy of $H_n \approx 0.6$ (calculated from the fractions) and the satisfaction of the $f_{\max,*} < 0.99$ condition confirm that the mixing and entropy priors have successfully suppressed the degenerate single channel attractors without enforcing a uniform distribution.

\subsection{Background evolution and numerical diagnostics}
\begin{figure}[htbp]
\centering
\includegraphics[width=0.45\textwidth]{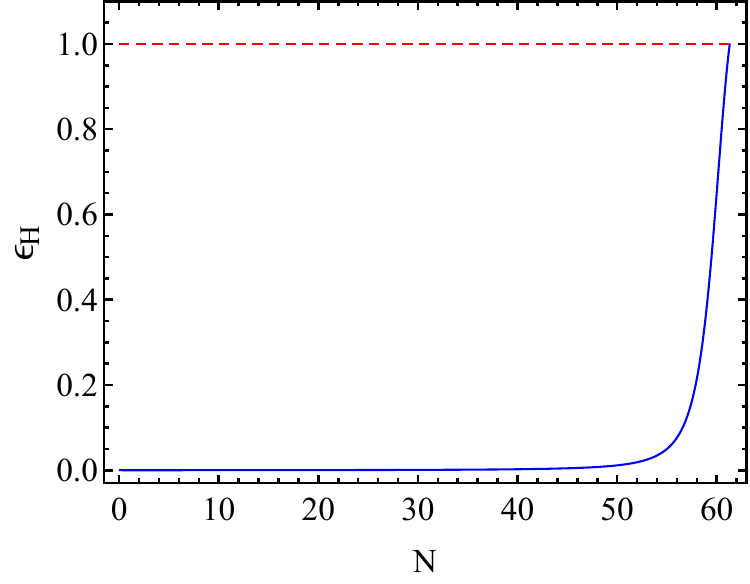}
\includegraphics[width=0.44\textwidth]{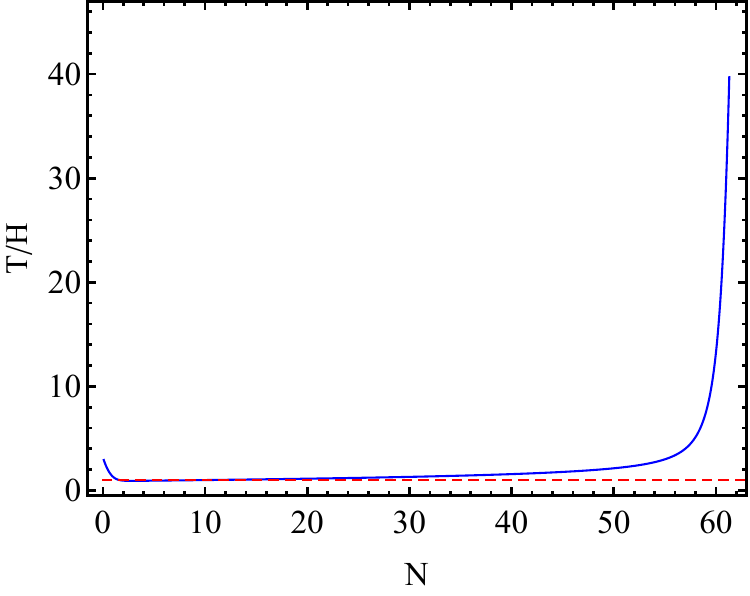}
\includegraphics[width=0.45\textwidth]{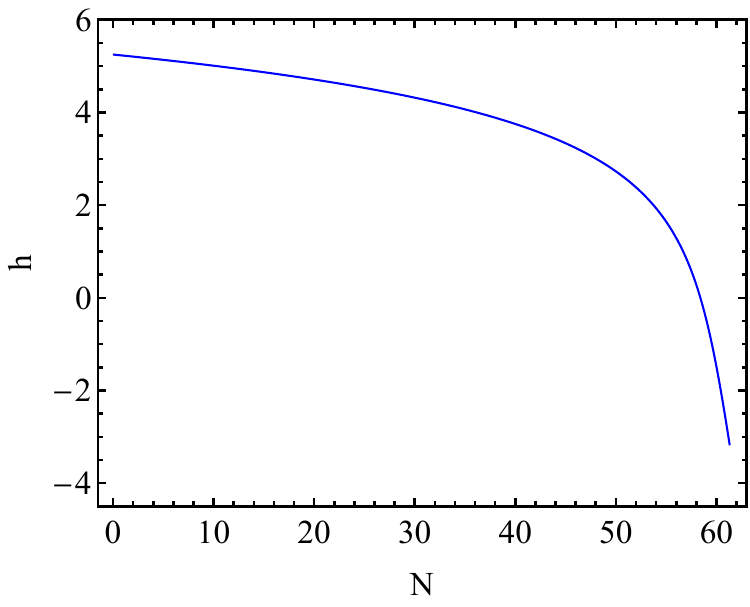}
\includegraphics[width=0.45\textwidth]{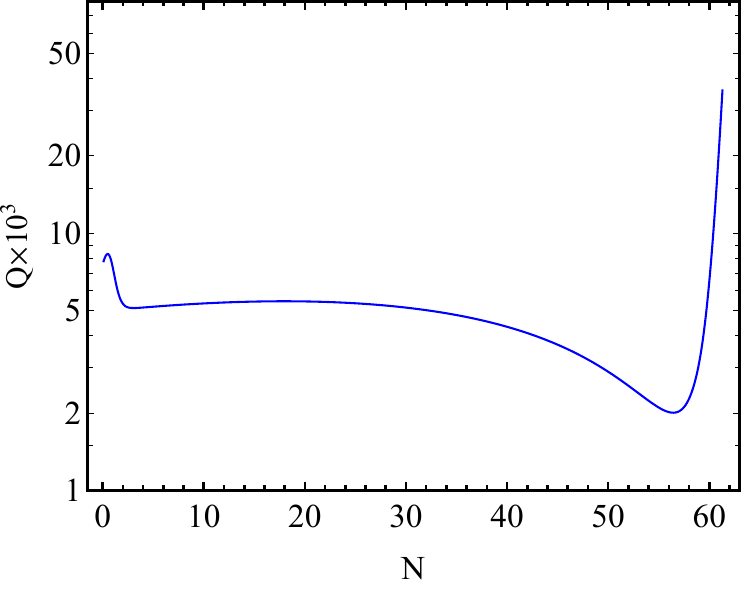}
\caption{Background evolution for the benchmark multi channel WHI solution, shown as a function of the number of e-folds $N$. The panels display: the Hubble slow-roll parameter $\epsilon_H(N)$ (top left), the temperature-to-Hubble ratio $T/H(N)$ (top right), the inflaton field trajectory $h(N)$ (bottom left), and the dissipation ratio $Q(N)$ (bottom right, scaled by $10^3$ for readability). Dashed lines indicate the $\epsilon_H=1$ and $T/H=1$ thresholds.}
\label{Fig:4N}
\end{figure}

Fig.~\ref{Fig:4N} displays the evolution of key background quantities as a function of the e-fold number $N$ for the benchmark solution.

The evolution of the Hubble slow-roll parameter, $\epsilon_H(N)$ (top left), exhibits behavior characteristic of plateau-like potentials. It remains negligible for the majority of the evolution before rising sharply to unity around $N \approx 60$, ensuring a clean and decisive end to inflation. This confirms that the background solution possesses a self-consistent termination mechanism.

The temperature-to-Hubble ratio, $T/H(N)$ (top right), remains close to the $T/H=1$ boundary for most of the inflationary period, indicating a "marginally warm" or "weakly warm" regime. Towards the end of inflation, however, the ratio surges to values significantly greater than one. This terminal heating phase signifies an efficient transfer of energy from the inflaton to the radiation bath as the system exits the slow-roll phase, a feature characteristic of a "graceful exit" in warm inflation models.

The inflaton field trajectory, $h(N)$ (bottom left), is observed to cross zero near the end of inflation. Since the effective potential and the dissipation coefficients depend only on $|h|$ (or $h^2$), the dynamics are invariant under the reflection $h \to -h$. This sign flip therefore carries no physical implication and is merely a consequence of the field's rapid motion in the steep, terminal region of the potential.

The dissipation ratio, $Q(N)$ (bottom right), remains small ($Q \ll 1$) throughout the entire inflationary epoch, indicating that the dynamics are always in the weakly dissipative regime. This is not in contradiction with the large values of $T/H$ observed at the end of inflation. In warm inflation dynamics, the rapid decrease of the Hubble parameter $H$ and the corresponding growth of the radiation energy density can jointly drive $T/H$ to large values even when $Q$ remains small.

\begin{figure}[htbp]
\centering
\includegraphics[width=0.6\textwidth]{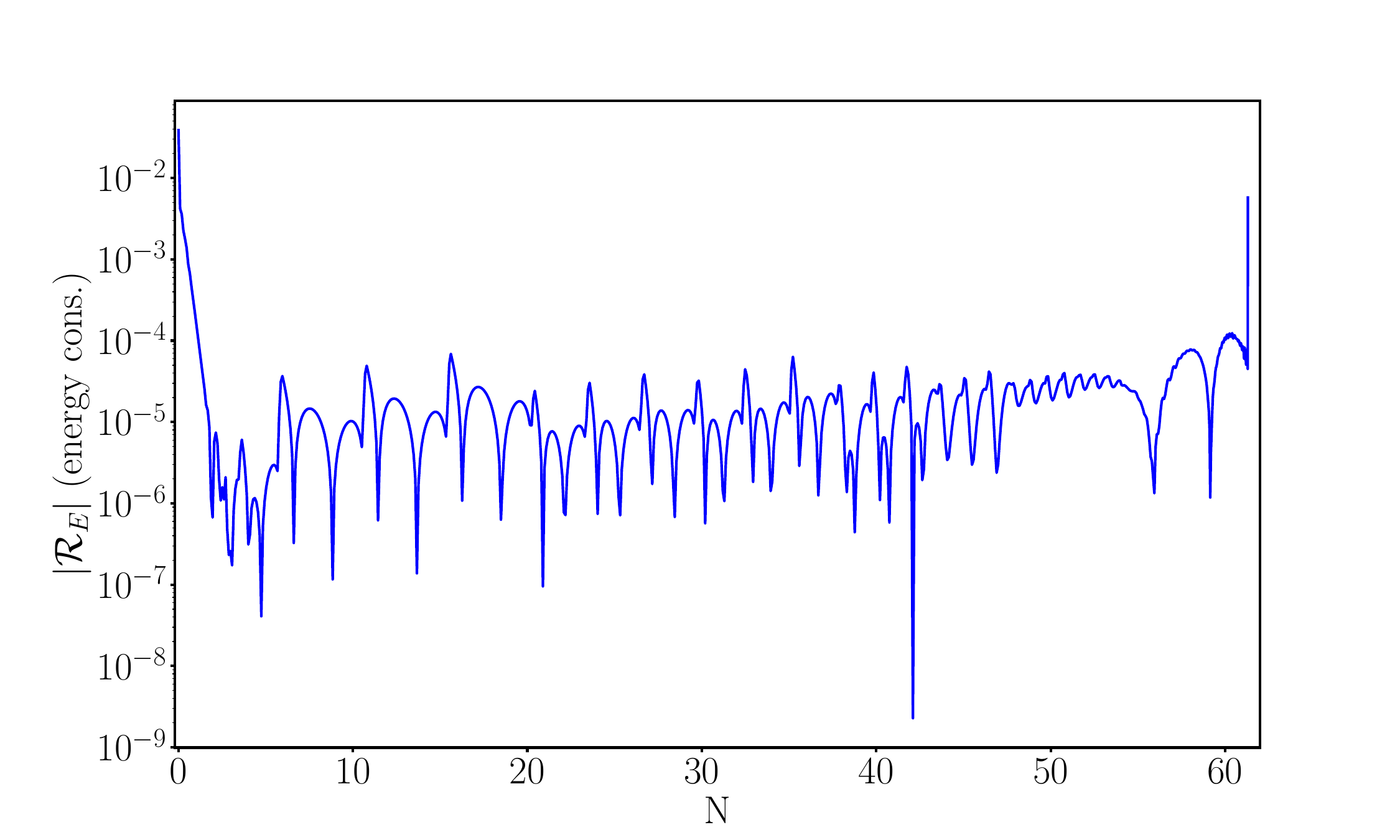}
\caption{Energy conservation diagnostic for the benchmark solution. The plot shows the absolute fractional energy balance residual, $|\mathcal{R}_E(N)|$, as a function of the number of e-folds $N$ (logarithmic vertical axis). The quantity $\mathcal{R}_E$ quantifies the instantaneous mismatch in the Friedmann equation as determined by the numerical integrator. Smaller values indicate better energy closure. The oscillatory pattern reflects the adaptive step-sizing of the integrator, while the sharp increase at the end corresponds to the increased stiffness of the system as it exits inflation.}
\label{Fig:E}
\end{figure}

Fig.~\ref{Fig:E} demonstrates the numerical stability of the solution. The energy conservation residual, $\mathcal{R}_E$, remains in the range of $10^{-7}$ to $10^{-4}$ throughout most of the inflationary epoch. Near the stiff exit phase, $\mathcal{R}_E$ transiently reaches a magnitude of $10^{-2}$. The diagnostic $\text{cons\_err}_{\max}$ is designed specifically to capture this peak value. The noticeable rise in the residual at the end of inflation is an expected numerical consequence of the system's dynamics becoming highly non-linear as $\epsilon_H \to 1$, and it remains within an acceptable range for a reliable solution.

\begin{figure}[htbp]
\centering
\includegraphics[width=0.475\textwidth]{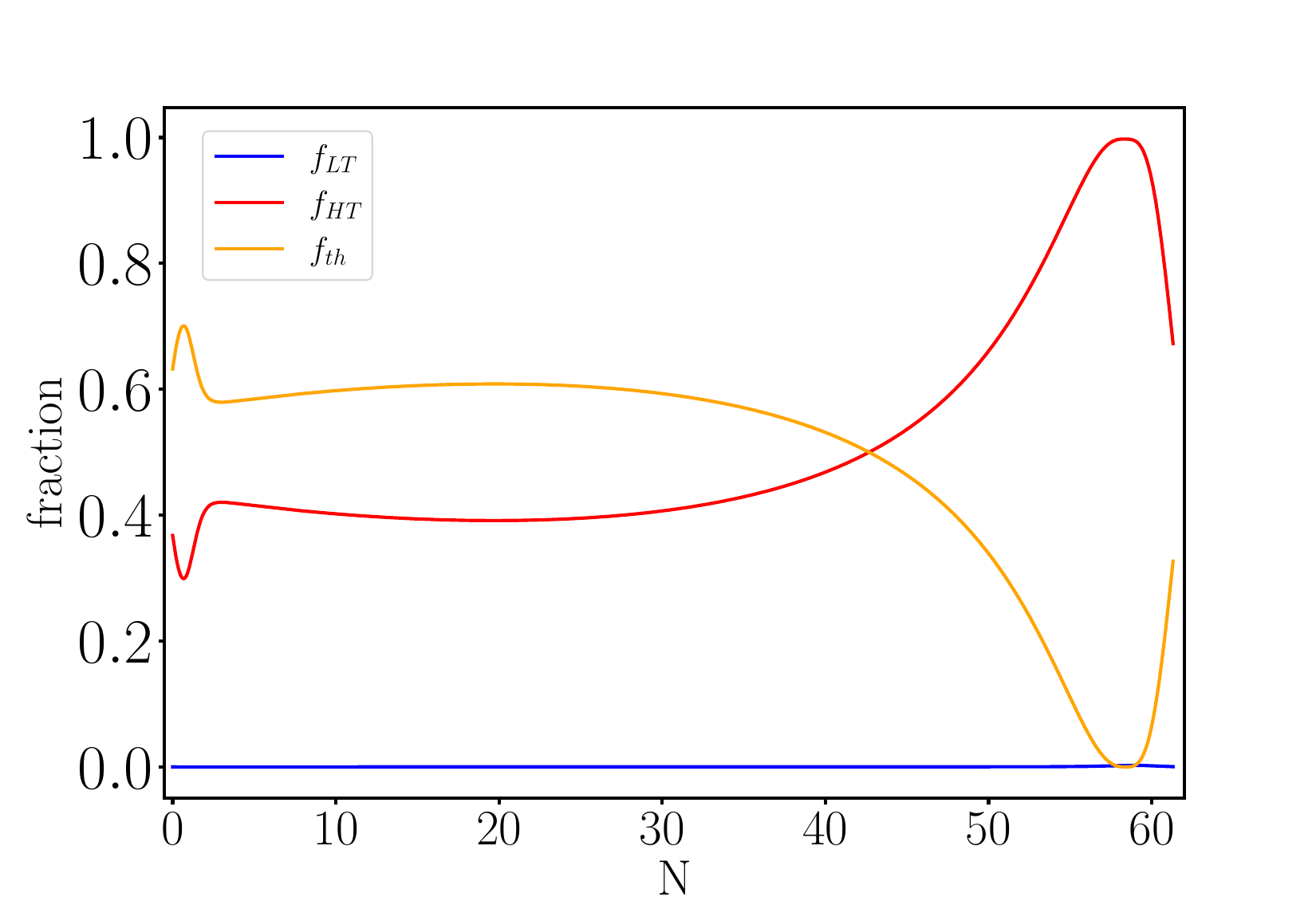}
\includegraphics[width=0.475\textwidth]{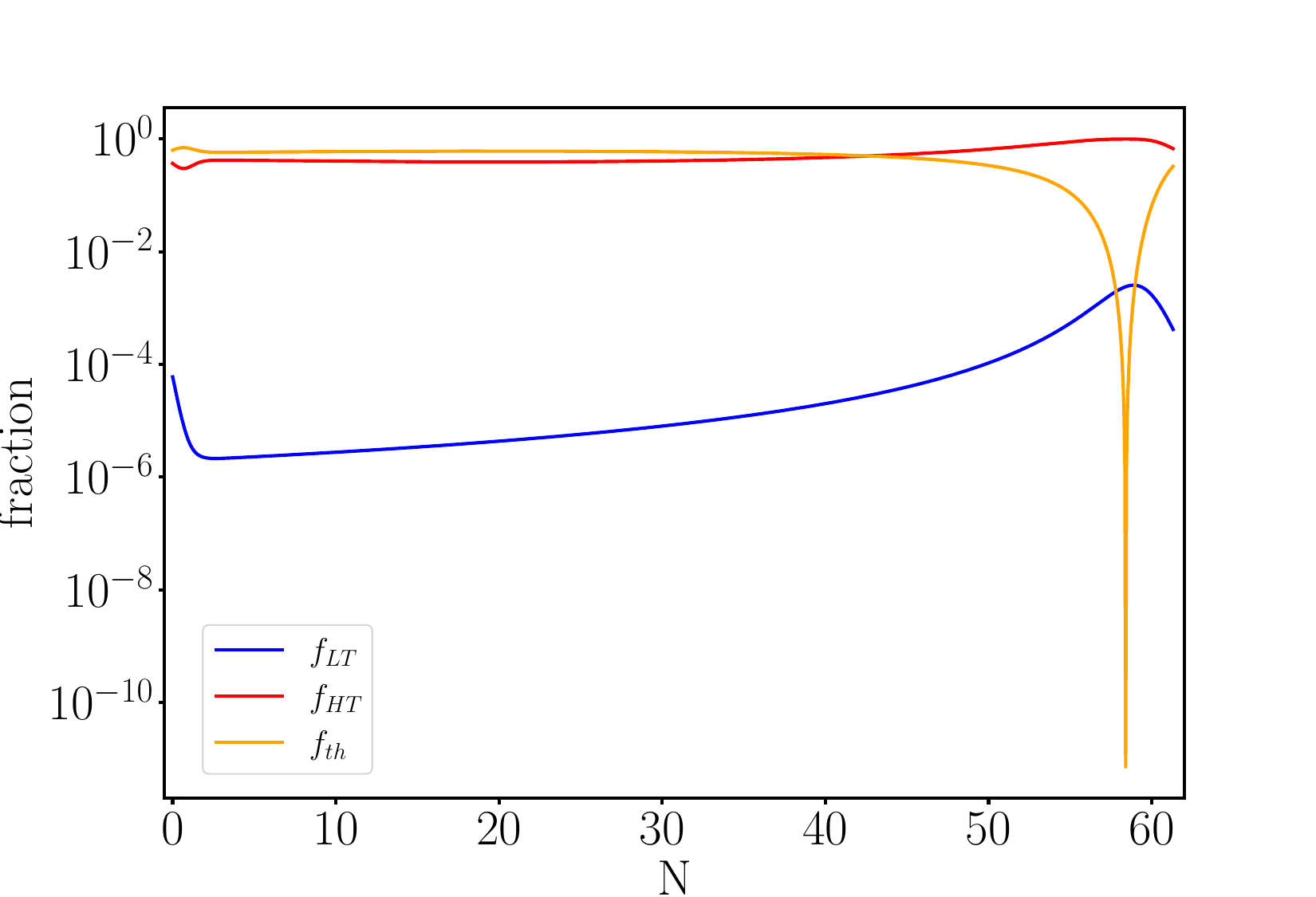}
\caption{Evolution of the fractional dissipation channel contributions, $f_i(N)$, along the background trajectory. The left panel uses a linear scale to emphasize the dominant partition, while the right panel uses a logarithmic scale to resolve subdominant components and diagnose sharp transitions.}
\label{Fig:fi}
\end{figure}

Fig.~\ref{Fig:fi} displays the evolution of the channel fractions, $f_i(N)$. For the majority of the trajectory, the total dissipation is genuinely shared between the HT and threshold channels, with $f_{\text{th}} \sim \mathcal{O}(0.6)$ and $f_{\text{HT}} \sim \mathcal{O}(0.4)$. The LT channel remains strongly suppressed, visible only on the logarithmic scale. As the system approaches the end of inflation ($N \to 61$), the channel composition exhibits a clear "relay" behavior: $f_{\text{HT}}$ increases rapidly to become dominant, while $f_{\text{th}}$ is sharply gated off by several orders of magnitude. This behavior is consistent with the Boltzmann suppression factor, $\exp[-M_{\text{eff}}(h,T)/T]$, in the definition of $\Upsilon_{\text{th}}$. This time-dependent reshuffling confirms that the solution does not represent a static, single channel regime. Instead, different dissipative mechanisms dominate at different epochs, which validates the motivation for the explicit three channel decomposition and the use of the structural diagnostic $f_i(N)$.

\subsection{Observable solution cloud in the $(n_s, r)$ plane}

\begin{figure}[h]
\centering
\includegraphics[width=1\textwidth]{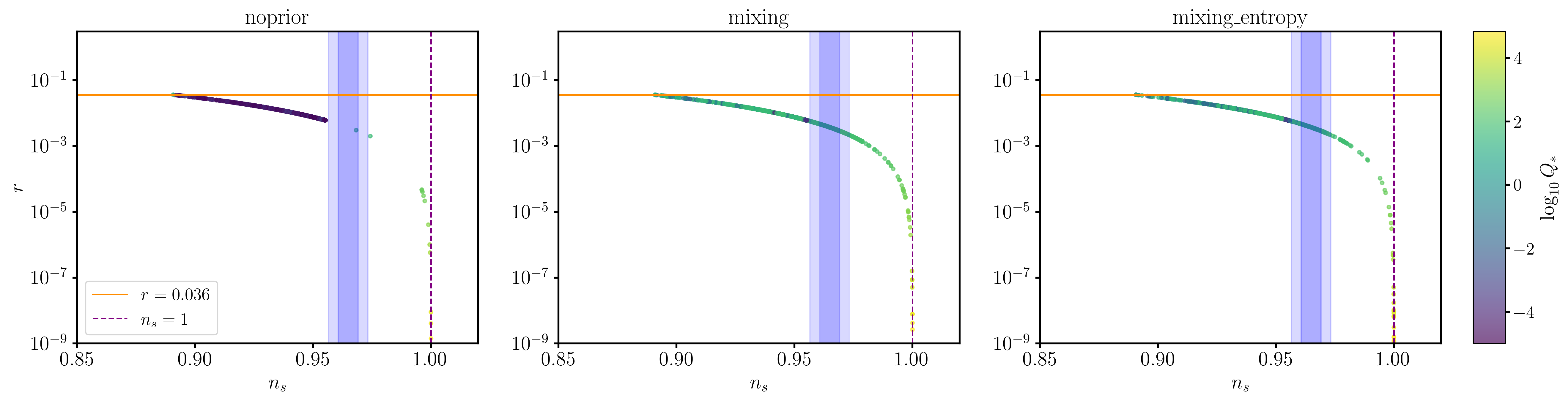}
\caption{Scatter plot of viable warm solutions in the $(n_s, r)$ plane for three inference settings: \texttt{noprior} (left), \texttt{mixing} (middle), and \texttt{mixing+entropy} (right). Each point corresponds to a solution passing the warm-selection criteria, colored by $\log_{10} Q_*$. The orange horizontal line denotes the reference upper bound $r = 0.036$~\cite{BICEP:2021xfz}, while the purple vertical dashed line marks the scale-invariant limit $n_s = 1$. The blue vertical bands indicate the Planck-allowed intervals for $n_s$~\cite{BICEP:2021xfz}.}
\label{Fig:nsr}
\end{figure}

Fig.~\ref{Fig:nsr} displays the distribution of warm-viable solutions in the $(n_s, r)$ plane. A robust qualitative feature across all three settings is the pronounced suppression of the tensor-to-scalar ratio, with solutions predominantly populating the small $r$ region. The point cloud exhibits a characteristic geometry: a smoothly connected "ridge" at $r \sim 10^{-3} \to 10^{-2}$ and an extension toward extremely small $r$ as $n_s$ approaches the scale invariant limit. This structure suggests that along the ridge, the effective slow-roll and dissipation parameters vary smoothly, producing a continuous family of solutions. The near vertical extension corresponds to a regime where the scalar sector is strongly enhanced relative to the tensor sector, driving $r$ to very small values. The color trend, encoded by $\log_{10}Q_*$, supports this interpretation: solutions with larger $Q_*$ preferentially occupy the very small $r$ region, consistent with the warm inflation mechanism where scalar fluctuations are enhanced by dissipation while tensor fluctuations are not.

A comparison of the three panels reveals that the global geometry of the $(n_s, r)$ manifold is largely preserved, while the population density of solutions is reshuffled by the structural priors. This separation of roles is physically informative. The spectral observables $(n_s, r)$ are primarily controlled by the background dynamics, whereas the priors act on the internal decomposition of the total dissipation, $\Upsilon = \sum_i \Upsilon_i$. By suppressing single channel dominated (low entropy) realizations, the priors improve the interpretability of the multi channel solutions without erasing features, such as the accumulation of points near the small-$r$ corner, which likely reflect a genuine large-volume region in the parameter space.

\begin{table}[htbp]
\caption{Counts of viable solutions under sequential selection cuts for the three inference settings.}
\begin{center}
\begin{tabular}{cccc}
\hline\hline
Setting & Warm & Warm + $(r<0.036)$ & Warm + $(r<0.036) + (n_s<1)$ \\
\hline
\texttt{noprior} & 14564 & 14485 & 14485 \\
\texttt{mixing} & 2148 & 1889 & 1889 \\
\texttt{mixing+entropy} & 2159 & 1971 & 1971 \\
Total & 18871 & 18345 & 18345 \\
\hline\hline
\end{tabular}
\end{center}
\label{Tab:prior}
\end{table}

Tab.~\ref{Tab:prior} quantifies the number of viable points shown in Fig.~\ref{Fig:nsr} under sequential selection cuts. The counts show that, under the warm selection alone, the \texttt{noprior} setting yields a substantially larger number of viable points. This indicates that a large fraction of the parameter space volume naturally collapses into channel-degenerate configurations. The structural priors significantly reduce this population by penalizing solutions with $f_{\max,*} \to 1$ and $H_n \to 0$, thereby reallocating the sample toward more interpretable multi channel realizations. The observational cuts on $r$ and $n_s$ remove only a modest fraction of points, confirming that the dominant effect of the priors is to control the channel-structure degeneracy, not the overall $(n_s, r)$ geometry.

\subsection{Effect of structural priors}

\begin{figure}[htbp]
\centering
\includegraphics[width=0.95\textwidth]{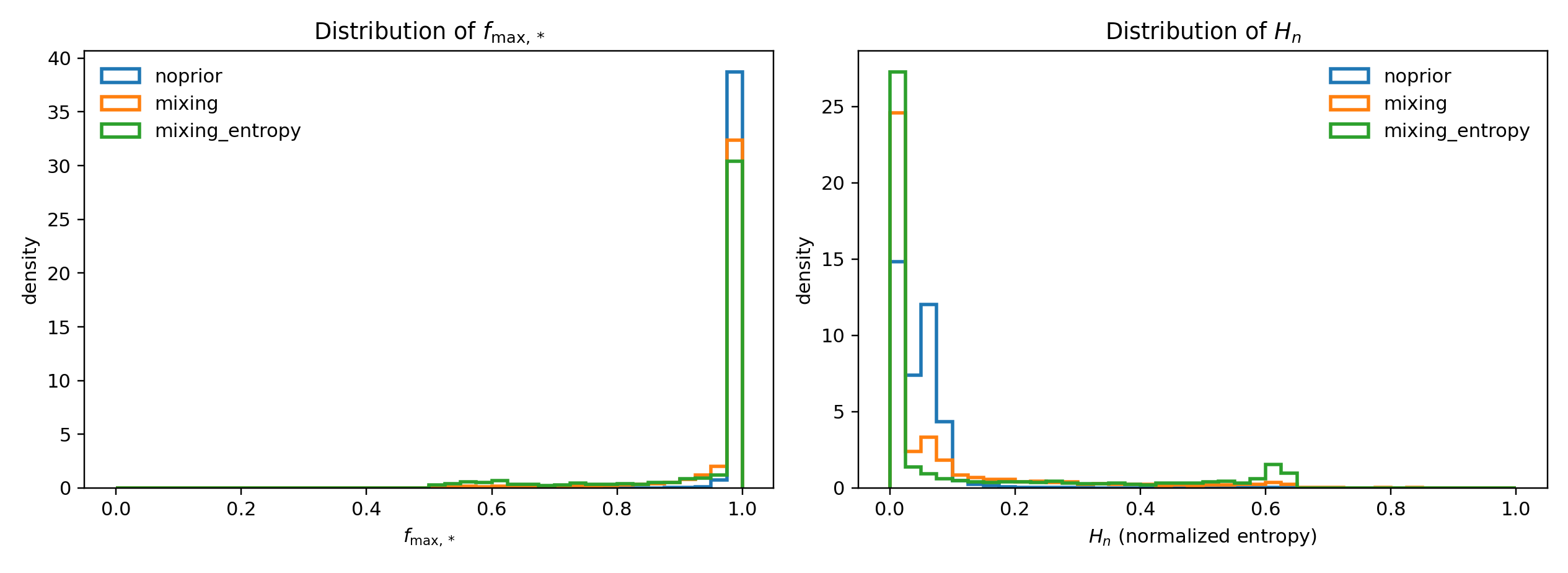}
\caption{Distributions of the channel dominance statistic $f_{\max,*}$ (left) and the normalized entropy $H_n$ (right) for warm viable solutions, comparing the three inference settings. These diagnostics provide complementary measures of degeneracy: $f_{\max,*} \to 1$ and $H_n \to 0$ indicate single channel dominance, while smaller $f_{\max,*}$ and larger $H_n$ indicate more balanced multi channel coexistence.}
\label{fig:fmHn}
\end{figure}

Fig.~\ref{fig:fmHn} reveals a strong tendency toward channel degeneracy in the raw parameter space. In the \texttt{noprior} setting, the distribution of $f_{\max,*}$ is sharply peaked at unity, while $H_n$ is concentrated near zero. This demonstrates that most viable solutions effectively collapse to a single channel description at the pivot scale, a consequence of large-volume regions in the parameter space where one channel's functional form overwhelms the others.

The introduction of the \texttt{mixing} prior partially redistributes the population away from this perfect dominance, broadening the $f_{\max,*}$ distribution and reducing the pile-up at $H_n \approx 0$. This forces the search algorithm to explore regions where at least two channels contribute non-negligibly, thereby improving the structural interpretability of the multi channel ansatz. The addition of the \texttt{mixing+entropy} prior further enhances this effect, creating a more pronounced tail at moderate entropies.

Nevertheless, a significant fraction of solutions remains near the degenerate limit even with priors. This persistence is physically plausible: a moderate-strength prior can reshape the population but cannot entirely erase a dominance peak that reflects a genuine large-volume effect in the underlying parameter space. The pair of statistics $(f_{\max,*}, H_n)$ thus provides a clean diagnostic of multi channel degeneracy, and the observed trend from \texttt{noprior} to \texttt{mixing+entropy} supports the central claim that structural priors enhance the interpretability of multi channel warm inflation solutions.

\section{Summary}\label{sec:Sum}

A three channel dissipative framework for Warm Higgs Inflation is constructed and analyzed. The total dissipation coefficient, $\Upsilon(h,T)$, is decomposed as $\Upsilon(h,T)=\Upsilon_{\rm LT}+\Upsilon_{\rm HT}+\Upsilon_{\rm th}$, which corresponds to the LT, HT, and th contributions, respectively. To achieve a global numerical solution and statistical inference of the background dynamics, a GA is employed. A key challenge in high-dimensional parameter scans is the ubiquitous single channel dominance degeneracy. To mitigate this, two structural priors are incorporated into the objective function: a \texttt{mixing} prior, which suppresses extreme pivot-scale channel fractions, and an \texttt{entropy} prior, which promotes multi channel coexistence through a larger normalized channel entropy. Furthermore, a layered warmness strategy, based on diagnostics such as $Q\equiv \Upsilon/(3H)$ and $T/H$, is utilized to decouple feasibility and model selection from the cosmological observables. This comprehensive workflow is demonstrated on a $14$-dimensional phenomenological model.

The numerical analysis is first anchored by a representative best-fit point, where the pivot observables $(n_s,r)$ and warmness indicators $(Q_*,T_*/H_*)$ are simultaneously controlled. The pivot-scale channel fractions for this point exhibit a genuinely mixed configuration (e.g., $f_{{\rm HT},*}\simeq 0.399480$, $f_{{\rm th},*}\simeq 0.600517$), confirming a multi channel realization rather than an effectively single channel scenario. The corresponding background evolution features a consistent slow-roll phase followed by a clean termination, with the slow-roll parameter $\epsilon_H$ rising toward unity near the exit. Warmness at the pivot is maintained by $T_*/H_*\gtrsim 1$, while the dissipation strength is typically weak ($Q_*\ll 1$) within the viable sample. The numerical reliability of the results is directly supported by the conservation diagnostic: the energy-conservation residual remains stably controlled at $\mathcal{O}(10^{-7} \to 10^{-4})$ throughout the evolution, including the exit stage where numerical stiffness may arise.

The time dependence of dissipation composition is clarified by the evolution of channel fractions $f_i(N)$, which reveals a pronounced ``channel relay'' along the trajectory. This behavior demonstrates that the effective dissipative mechanism can change during a single inflationary history, providing a concrete interpretation of mixed pivot states. Consequently, the explicit three channel decomposition is motivated as a crucial diagnostic for interpretability and inference. At the population level, the $(n_s,r)$ solution clouds for the \texttt{noprior}, \texttt{mixing}, and \texttt{mixing+entropy} cases exhibit a broadly stable global geometry. This indicates that the observable manifold is primarily controlled by the background dynamics and the effective dissipation strength, rather than by the detailed channel partition. In contrast, the viability counts and structural statistics clearly show that the structural priors substantially reshape the solution population by suppressing channel-degenerate realizations. An ablation study yields $18871$ viable points; after imposing the constraint $r<0.036$, the retained numbers become $14485$, $1889$, and $1971$ for the \texttt{noprior}, \texttt{mixing}, and \texttt{mixing+entropy} cases, respectively. Collectively, these results establish a numerically reliable and statistically tractable pipeline for generating and diagnosing multi channel warm inflation backgrounds, providing a basis for subsequent precision computations of primordial perturbation spectra on selected representative solutions.

\hspace{2cm}
\acknowledgments
Wei Cheng was supported by Chongqing Natural Science Foundation project under Grant No. CSTB2022NSCQ-MSX0432, by Science and Technology Research Project of Chongqing Education Commission under Grant No. KJQN202200621, and by Chongqing Human Resources and Social Security Administration Program under Grants No. D63012022005.

%\bibliography{lit}

\bibliographystyle{arxivref}

\appendix

\section{Appendix A: Full parameter ranges and priors used in the GA scan}

Full scan ranges for the 14 free parameters in the three channel WHI setup are presented in Tab.~\ref{tab:param_bounds}. Parameters denoted with $\log_{10}(\cdot)$ are sampled uniformly in log space (log-uniform in linear space), while the remaining parameters are sampled uniformly in linear space. All parameters are dimensionless in reduced Planck units: $h_0$ (in this table) denotes the initial field value at the start of integration, and $\mu$, $m_0$ set the characteristic mass scales associated with threshold activation and thermal corrections.

\begin{table}[htb]
\caption{Full scan ranges for the $14$ free parameters in the three channel WHI setup.
Parameters listed as $\log_{10}(\cdot)$ are sampled uniformly in $\log_{10}$ space (equivalently log-uniform in linear space),
while the remaining parameters are sampled uniformly in the linear variable unless stated otherwise.
All parameters are dimensionless in reduced Planck units; $h_0$ denotes the initial field value at the start of integration,
and $\mu$ and $m_0$ set the characteristic mass scales entering the threshold activation and thermal corrections.}
\begin{center}
\begin{tabular}{c c c c}
\hline\hline
parameter & range & parameter & range \\
\hline
$\lambda$ & $[10^{-6},\ 10^{-1}]$ & $c$ & $[-1.0,\ 6.0]$ \\
$\xi$ & $[1.0,\ 10^6]$ & $\alpha_h$ & $[-0.8,\ 0.8]$ \\
$\log_{10} C_{\text{LT}}$ & $[-12.0,\ 6.0]$ & $\alpha_T$ & $[-0.8,\ 0.8]$ \\
$\log_{10} C_{\text{HT}}$ & $[-12.0,\ 6.0]$ & $\log_{10} h_0$ & $[-4.0,\ 1.0]$\\
$\log_{10} C_{\text{th}}$ & $[-12.0,\ 6.0]$ &$\log_{10} \mu$ & $[-3.0,\ 1.0]$\\
$g$ & $[10^{-4},\ 1.5]$   & $\log_{10} m_0$ & $[-3.0,\ 2.0]$ \\
$a$ & $[-4.0,\ 4.0]$ & $\kappa_T$ & $[0.0,\ 20.0]$  \\
\hline\hline
\end{tabular}
\end{center}
\label{tab:param_bounds}
\end{table}

\section{Appendix B: GA Configuration Parameters}

Configuration of the GA employed in this work is summarized in Tab.~\ref{tab:GA_detailed_config}. The upper block lists core hyperparameters (e.g., population size, number of generations) that determine the evaluation budget, the lower block (in this table) specifies genetic operators and mechanisms (tournament selection, crossover/mutation parameters, elitism rule) that define the algorithm's execution logic.

A more detailed explanation is as follows:
\begin{itemize}
\item \textbf{Hyperparameters}
\begin{itemize}
    \item \textbf{pop\_size (Population size)}: The total number of individuals participating in each generation of evolution. It determines the search breadth of the algorithm: a larger value helps avoid missing optimal solutions but increases computational cost.

    \item \textbf{n\_generations (Number of generations)}: The total number of iterative evolution cycles of the GA. Sufficient iterations ensure the algorithm has enough time to converge toward the global optimal solution.
    \item \textbf{cx\_prob (Crossover probability)}: The probability of gene crossover/recombination between two individuals in the population. It promotes the fusion of excellent gene segments and expands the search range of the solution space.
    \item \textbf{mut\_rate (Mutation probability)}: The probability of random mutation in an individual’s genes. It helps the algorithm avoid falling into local optimal solutions and maintains population diversity.
    \item \textbf{mut\_scale (Mutation scale)}: The numerical adjustment range when gene mutation occurs. It controls mutation intensity: an excessively large scale may destroy excellent genes, while an excessively small scale limits search ability.
    \item \textbf{seed (Random seed)}: Fixes the initial state of the random number generator, ensuring consistent results across multiple runs of the algorithm (critical for experimental reproducibility).
\end{itemize}
\end{itemize}

\begin{itemize}
\item \textbf{Operators \& Mechanisms}
\begin{itemize}
    \item \textbf{Selection Operator (Type: tournament)}: A selection method that randomly selects a subset of individuals and picks the best one from the subset. It balances selection pressure (promoting excellent individuals) and population diversity (avoiding premature convergence).
    \item \textbf{Crossover Operator (Probability}: The probability of gene crossover between individuals (consistent with $cx\_prob$). It is applied in the crossover function to combine genes from different individuals.
    \item \textbf{Mutation Operator (Rate)}: The probability of gene mutation (consistent with $mut\_rate$). It is applied in the mutate function to introduce random genetic variations.
    \item \textbf{Mutation Operator (Scale)}: The numerical range of gene mutation (consistent with $mut\_scale$). It is applied in the mutate function to control the intensity of genetic adjustments.
    \item \textbf{Elitism Mechanism (Rule)}: A rule that retains the top-performing individuals from the current generation to the next. It ensures excellent genetic information is not lost during evolution.
\end{itemize}
\end{itemize}

\begin{table}[htb]
\caption{GA configuration used in this work. The upper block lists the core hyperparameters that determine the evaluation budget, while the lower block specifies the genetic operators and mechanisms adopted in our implementation, including tournament selection, crossover/mutation settings, and the elitism rule that retains the top-performing individuals each generation.}
\begin{center}
\begin{tabular}{c c c c}
\hline\hline
GA Component & Type/Parameter & Value & Description \\
\hline
\multicolumn{4}{c}{Hyperparameters} \\
\cline{1-4}
- & pop\_size & 180 & Population size \\
- & n\_generations & 160 & Number of generations \\
- & cx\_prob & 0.5 & Crossover probability \\
- & mut\_rate & 0.25 & Mutation probability \\
- & mut\_scale & 0.18 & Mutation scale \\
- & seed & 42 & Random seed \\
\hline
\multicolumn{4}{c}{Operators \& Mechanisms} \\
\cline{1-4}
Selection Operator & Type & tournament & Tournament selection \\
Crossover Operator & Probability & 0.5 & Used in crossover function \\
Mutation Operator & Rate & 0.25 & Used in mutate function \\
 & Scale & 0.18 & Used in mutate function \\
Elitism Mechanism & Rule & $\max(1, pop\_size//12)$ & Retain top individuals \\
\hline\hline
\end{tabular}
\end{center}
\label{tab:GA_detailed_config}
\end{table}

\section{Appendix C: Default hyperparameters}

The default hyperparameter configuration adopted in this paper is summarized in Tab.~\ref{tab:hyperparams}, which includes key parameters related to model training and constraint conditions (e.g., the dominance control parameter $f_{\text{dom}}$ and the minimum contribution of the threshold channel $f_{\text{th,min}}$). The values of these parameters are determined based on a comprehensive consideration of numerical stability and physical plausibility.

\begin{table}[htb]
\caption{Default hyperparameters (as used in the implementation).}
\label{tab:hyperparams}
\begin{center}
\begin{tabular}{c c c c}
\hline\hline
hyperparameter & value & hyperparameter & value \\
\hline
$n_s^{\rm obs}$ & $0.9649$ & $\sigma_{n_s}$ & $0.0042$ \\
$r_{\max}$ & $0.036$ & $w_r$ & $100$ \\
$(T/H)_{\min}$ & $1.6$ & $Q_{\min}$ & $10^{-5}$ \\
$W_{T/H}^{(0)}$ & $50$ & $W_Q^{(0)}$ & $50$ \\
$A_{\rm floor}$ & $0.2$ & $p$ & $2.0$ \\
$f_{\rm dom}$ & $0.99$ & $W_{\rm dom}$ & $10.0$ \\
$f_{{\rm th,min}}$ & $10^{-3}$ & $W_{\rm th}$ & $0.01$ \\
$H_{\min}$ & $0.75$ & $W_{\rm ent}$ & $20.0$ \\
$f_{{\rm each,min}}$ & $5\times 10^{-4}$ & $W_{\rm each}$ & $0.02$ \\
$N_{\max}$ & $120$ & $w_N$ & $10^{-3}$ \\
$\mathcal{R}_{E,{\rm thr}}$ & $10^{-3}$ & $w_{\rm cons}$ & $50.0$ \\
$\chi^2_{\rm fail}$ & $10^{6}$ & $\chi^2_{\rm hard}$ & $10^{6}$ \\
\hline\hline
\end{tabular}
\end{center}
\end{table}

\end{document}